\documentclass[twocolumn]{aastex631}
\usepackage{comment}
\usepackage{color}

\newcommand\m{\mathrm}
\newcommand{\twelveco}{\mbox{$^{12}$CO}} 
\newcommand{\thirteenco}{\mbox{$^{13}$CO}} 
\newcommand{\ceighteeno}{\mbox{C$^{18}$O}} 
\newcommand{\twelvecol}{\mbox{$^{12}$CO\,({\it J}=1--0})} 
\newcommand{\thirteencol}{\mbox{$^{13}$CO\,({\it J}=1--0})}

\newcommand{\ceighteenol}{\mbox{C$^{18}$O({\it J}=1--0)}}

\newcommand{\hcopl}{\mbox{HCO$^{+}$({\it J}=1--0)}} 
 
\newcommand{\hcnl}{\mbox{HCN({\it J}=1--0)}}

\newcommand{\siol}{\mbox{SiO({\it J}=2--1)}}

\newcommand {\kms}{\mbox{km~s$^{-1}$}}

\newcommand {\vlsr}{\mbox{$V_{\rm LSR}$ }}

\newcommand {\tastar}{\mbox{$T_\mathrm{a}^{*}$}}

\usepackage{xcolor}
\usepackage{ulem}

\begin{document}


\title{
Progenitor Constraint Incorporating Shell Merger: The Case of Supernova Remnant G359.0$-$0.9
}

\author[0009-0003-0653-2913]{Kai Matsunaga}
\affiliation{Department of Physics, Graduate School of Science, Kyoto University, Kitashirakawa Oiwake-cho, Sakyo-ku, Kyoto 606-8502, Japan}

\author[0000-0003-1518-2188]{Hiroyuki Uchida}
\affiliation{Department of Physics, Graduate School of Science, Kyoto University, Kitashirakawa Oiwake-cho, Sakyo-ku, Kyoto 606-8502, Japan}

\author[0000-0003-2735-3239]{Rei Enokiya}
\affiliation{Faculty of Science and Engineering, Kyushu Sangyo University, 2-3-1 Matsukadai, Fukuoka 813-8503, Japan}

\author[0000-0001-9267-1693]{Toshiki Sato}
\affiliation{Department of Physics, School of Science and Technology, Meiji University, 1-1-1 Higashi Mita, Tama-ku, Kawasaki, Kanagawa 214-8571, Japan}

\author[0000-0003-4876-5996]{Ryo Sawada}
\affiliation{Department of Earth Science and Astronomy, Graduate School of Arts and Sciences, The University of Tokyo, Tokyo 153-8902, Japan}

\author[0000-0001-8338-502X]{Hideyuki Umeda}
\affiliation{1Department of Astronomy, School of Science, University of Tokyo, Tokyo 113-0033, Japan}

\author[0009-0006-7889-6144]{Takuto Narita}
\affiliation{Department of Physics, Graduate School of Science, Kyoto University, Kitashirakawa Oiwake-cho, Sakyo-ku, Kyoto 606-8502, Japan}

\author[0000-0002-5504-4903]{Takeshi Go Tsuru}
\affiliation{Department of Physics, Graduate School of Science, Kyoto University, Kitashirakawa Oiwake-cho, Sakyo-ku, Kyoto 606-8502, Japan}

\begin{abstract}
It is generally hard to put robust constraints on progenitor masses of supernovae (SNe) and remnants (SNRs) observationally, while they offer tantalizing  clues to understanding explosion mechanisms and mass distribution.
Our recent study suggests that ``shell merger'', which is theoretically expected for stellar evolution, can  appreciably affect final yields of inter-mediate mass elements (IMEs; such as Ne, Mg, and Si).
In light of this, here we report results of  X-ray spectral analysis of a Galactic SNR G359.0$-$0.9, whose abundance pattern may possibly be anomalous according to a previous study.
Our spectroscopy using all the available data taken with XMM-Newton reveals that this remnant is classified as Mg-rich SNRs because of its high Mg-to-Ne ratio ($\m{Z_{\rm Mg}/Z_{\rm Ne}}=1.90^{+0.27}_{-0.19}$; mass ratio $0.66^{+0.09}_{-0.07}$) and conclude that the result cannot be explained without the shell merger.
By comparing the observation with theoretical calculations, we prefer the so-called  Ne-burning shell intrusion and in this case the progenitor mass $M_\m{ZAMS}$ is likely  $<15M_\odot$.
We confirm the result also by our new molecular line observations with the NRO-45~m telescope: G359.0$-$0.9 is located in the Scutum-Centaurus arm (2.66--2.94~kpc) and in this case the resultant total ejecta mass  $\sim6.8M_\odot$ is indeed consistent with the above estimate.
Our method using mass ratios of IMEs  presented in this paper will become useful to distinguish the type of the shell merger, the Ne-burning shell intrusion and the O-burning shell merger, for future SNR studies.

\end{abstract}

\section{Introduction}
Core-collapse (CC) supernovae (SNe)  spread heavy elements produced  during their progenitor's evolution and ensuing explosive end, and thus abundances of ejecta in supernova remnants (SNRs) should provide rich information on  their past stellar and explosive nucleosynthesis.
Although a wide variety of elemental compositions has been reported so far from X-ray spectroscopy of SNRs, a plausible connection between their abundances and  progenitor properties is still missing.
This is partially because the yield of nucleosynthesis products depends on explosion  mechanisms  which are currently not fully understood.
$\m{^{56}Ni}$ (decaying to $\m{^{56}Fe}$) is, for instance, created near the proto-neutron star (PNS), where there is a large theoretical uncertainty.
While, as \citet{Katsuda_2018} reported, the abundance ratio of Fe/Si is known as one of the promising parameters for estimating zero age main sequence mass ($M_\m{ZAMS}$), a total mass of $\m{^{56}Ni}$ varies with different explosion engines  \citep[e.g.,][and references therein]{sawada_2019, ertl_2020, imasheva_2023, sawada_2023}.


Lighter elements such as O, Ne, and Mg are mainly synthesized in the pre-SN phase, which may give us more direct hints in understanding progenitors.
In our recent study \citep{sato_2024a}, we suggested that the high Mg/Ne ratio in the ejecta of a SNR N49B can be explained by a destratification process, in agreement with progenitor models published by \citet{sukhbold_2018}.
Such so-called "shell merger" processes are also reported by recent multi-dimension simulations of the progenitor interior \citep[c.f.][]{mocak_2018,andrassy_2020,yadav_2020}.
According to \citet{collins_2018} and our study \citep{sato_2024a} with the models of \citet{sukhbold_2018}, such events will commonly occur in several tenths of a percent of progenitors, which may also affect the chemical evolution of galaxies \citep{ritter_2018}.
As indicated by \citet{yadav_2020}, when the O- or Ne-burning shell is merged with the outer O-Ne-Mg layer before the collapse, the total yield of Ne is heavily depleted and the resultant Mg/Ne will become higher.
We also found that if such violent mergers of burning shells occur the compactness parameter \citep{oconnor_2011} tends to become smaller \citep{sato_2024a}:  with this stellar evolution path even a supermassive star with $M_\m{ZAMS}\gtrsim20M_\odot$ may become explodable. 
Mg-rich SNRs are therefore possibly massive-star remnants while observations of SNe show no robust evidence for an explosion of a red supergiant whose $M_\m{ZAMS}$ exceeds $20M_\odot$ \citep[red supergiant problem;][and references therein]{davies_2020}.

Although only a few Mg-rich SNRs has been reported so far \citep[N49B; G284.3-1.8,][]{park_2017, williams_2015}, more candidates will be identified  by precisely measuring the Ne and Mg abundances; most promising targets  for this purpose would be low-absorbed, i.e., nearby remnants.
Here, we focus on  the analysis of G359.0$-$0.9, which has a large apparent size, suggesting a nearby SNR.
G359.0$-$0.9 was first discovered by a 10-GHz radio continuum survey with the Nobeyama 45-m (NRO 45-m) telescope \citep{Sofue_1984} and follow-up observations revealed a partial radio shell in 1.2~GHz \citep[MeerKAT;][]{heywood_2022} and H$\alpha$ \citep[Anglo-Australian Observatory/United Kingdom Schmidt Telescope;][]{stupar_2011}.
X-ray observations have been performed with ASCA \citep{bamba_2000} and Suzaku \citep{bamba_2009}, which have revealed an extremely asymmetric morphology and also suggested a relatively large Mg abundance of ejecta.
Although these previous studies imply that G359.0$-$0.9 is another candidate for massive-star remnants, the ejecta mass, which correlates with $M_\m{ZAMS}$ \citep[e.g.,][]{sukhbold_2016},  is still uncertain due to discrepant distance estimates; $\sim6~\m{kpc}$ \citep[X-ray absorption;][]{bamba_2000}, $3.5\pm{0.4}~\m{kpc}$  or $3.3\pm{0.2}~\m{kpc}$ \citep[luminosity attenuation of red clumps;][]{wang_2020}, and $\sim3.7~\m{kpc}$ \citep[$\Sigma$-$D$ relation;][]{pavlovic_2012}.

In this paper, we present the results of X-ray and molecular line observations of G359.0$-$0.9 with XMM-Newton and NRO 45-m.
This paper is organized as follows.
Section~\ref{sec:obs} details the above observations and data reduction method.
In Section~\ref{sec:ana}, we show results of spectroscopy and provide evidence that this remnant is classified as Mg-rich SNRs.
We also report discovery of a molecular cloud associated with G359.0$-$0.9.
We then discuss the origin of the Mg-rich SNRs in Section~\ref{sec:dis} and infer the progenitor mass of G359.0$-$0.9 in conjunction with a distance estimation 
through the velocity of the associated cloud.
Throughout the paper, errors of parameters are defined as 1$\sigma$ confidence level and the solar abundance is given by \citet{wilms_2000}.





\section{Observation and data reduction}\label{sec:obs}
\subsection{X-ray observation with XMM-Newton}

\begin{table*}[t]
    \centering
    \caption{Observation log of G359.0$-$0.9 with XMM-Newton.}
    \begin{tabular}{ccccc}
    \hline
       Obs. ID & Date of Obs. & Detector & Total Exposure (ks) & Effective Exposure (ks)\\
    \hline\hline
        0152920101 & 2003-04-02 & MOS1 & 51.5 & 48.1\\
                    &       & MOS2 & 51.5 & 47.0\\
                    &         & pn    & 48.9 & 36.8\\
        0801680501 & 2017-09-18 & MOS1 & 26.9 & 20.4\\
                    &           & MOS2 & 28.1 & 22.2\\
                    &           & pn    & 22.9 & 16.0\\
        0801680701 & 2017-09-23 & MOS1 & 28.5 & 26.9\\
                    &           & MOS2 & 28.6 & 27.1\\
                    &           & pn & 26.1 & 23.6\\
        0804250301 & 2018-03-13 & MOS1 & 40.6 & 39.5\\
                    &           & MOS2 & 40.6 & 39.7\\
                    &           & pn   & 39.5 & 38.4\\
    \hline
    \end{tabular}
    \label{tab:obs}
\end{table*}


G359.0$-$0.9 has been observed with XMM-Newton four times so far as summarized in Table~\ref{tab:obs}.
For the following analysis, we used  data from European Photon Imaging Camera (EPIC) instruments: two metal-oxide semiconductor (MOS) detectors \citep{turner_2001} and one pn detector \citep{struder_2001}. 
Data reduction was performed using  the XMM Science Analysis System (SAS) version 19.1.0 with the most recent Current Calibration Files (CCF).
We combined the spectral files using \texttt{mathpha} from the HEASoft software package.
The redistribution matrix file (RMF) and ancillary response file (ARF) for each observation were multiplied using \texttt{marfrmf}.
They were then merged using \texttt{addrmf}, weighting by their effective exposure times and detector areas.
For the following spectral analysis, we used version 12.13.0c of an X-Ray Spectral Fitting Package \citep[XSPEC,][]{arnaud_1996}.
 

\subsection{Nobeyama 45-m molecular line observations}\label{sec:nobeyama}
Molecular line observations were carried out by two separated sessions in April 2023 by using the Nobeyama 45~m telescope (PI; K. Matsunaga).
We used FOur-beam REceiver System on the 45-m Telescope \citep[FOREST; ][]{min16}, and Spectral Analysis Machine for the 45-m telescope \citep[SAM45;][]{kam12} in the frontend and backend, respectively.
FOREST is comprised of four beams with two polarization and two sidebands. 
Thanks to the large intermediate frequency (IF) bandwidths of 8~GHz (4--12~GHz), a simultaneous observation with $\twelvecol$, $\thirteencol$, and $\ceighteenol$ is realized.
A typical system noise temperature toward the source and the image rejection ratio were $\sim$500~K, and 15~dB, respectively.
SAM45 is a FX-type correlator, which consists of 16 IF bands (4 beams$\times$2 polarizations$\times$2 sidebands) with a 2-GHz bandwidth.
We configured two settings for the spectrometer: CO isotopes  in a first wide-area survey (setting~1), and follow-up, small-area observations with HCO+, SiO, and HCN (setting~2).
We used a ``spectral window mode'', which produces two independent spectral windows within each IF band.
The roles of 32 spectral windows to the two spectrometer settings are summarized in Table~\ref{tab:setting}.
The narrow/wide bands correspond to the velocity resolutions and velocity coverages of 0.33/0.67~$\kms$ and $\pm$300/$\pm$600~$\kms$, respectively.

\begin{table*}[t]
    \centering
    \caption{Roles of spectral windows by settings.}
    \begin{tabular}{ccccc}
    \hline
       Setting & A01 to A08 & A09 to A16 & A17 to A24 & A25 to A32\\
    \hline\hline
        1 & $\twelvecol$ wide band & $\ceighteenol$ narrow band & $\twelvecol$ narrow band & $\thirteencol$ narrow band\\
       2 & $\hcopl$ wide band & $\siol$ wide band & $\hcnl$ wide band & $\siol$ narrow band\\
    \hline
    \end{tabular}
    \label{tab:setting}
\end{table*}

Through a deep position-switching observation toward ($l$, $b$) = ($-$1\fdg1000, $-$1\fdg8000) by beam 1, we found that this direction does not include any significant emissions over a 3$\sigma$ noise level in all beams, and thus we used this coordinate as the emission-free direction to do a chopper wheel calibration \citep{kut81}.
Pointing observations were done every two hours by observing SiO masers (OH2.6$-$0.4) with the H40 receiver at 40~GHz, and achieved a pointing accuracy better than 2\arcsec.
We carried out mapping observations of the large area shown in Figure~\ref{LBch} with setting~1, and limited areas with setting~2.
The half-power beam widths (HPBW), velocity resolutions, and typical r.m.s. noise fluctuations of the final datasets are $\sim$26$\arcsec/$26$\arcsec$/26$\arcsec$, 0.25/0.25/0.25~$\kms$, and 0.60/0.17/0.17~K per channel in $\tastar$ scale for 
$\twelveco$, $\thirteenco$, and $\ceighteeno$, respectively.
$\siol$ emissions were not detected in our observations.

\section{Data Analysis}\label{sec:ana}
\begin{figure}[t]
    \centering
    \includegraphics[width=\linewidth]{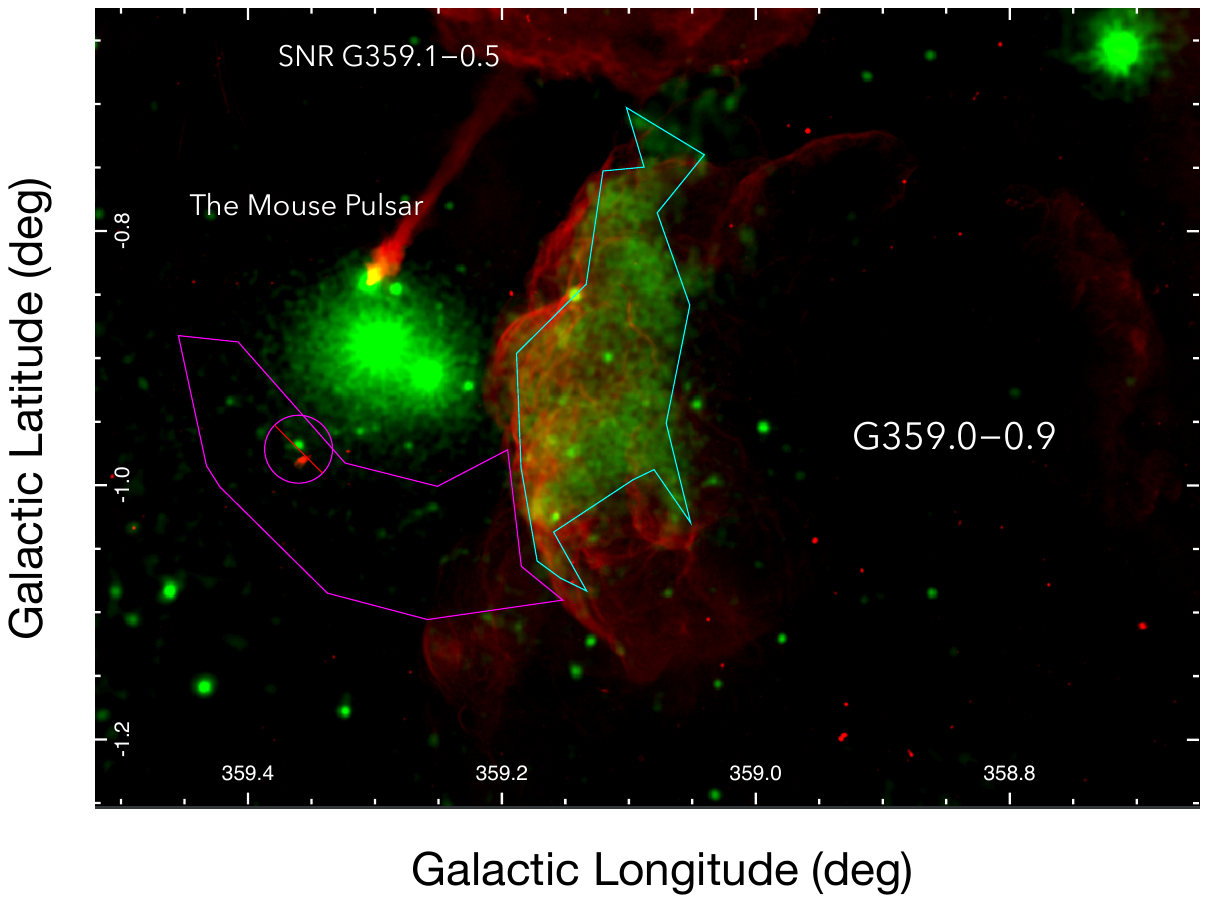}
    \caption{
    X-ray image of G359.0$-$0.9 (1.0--2.0~keV) and the surrounding field obtained with XMM-Newton (green).
    The radio continuum emission obtained with MeerKAT at 1284 MHz is overlaid in red.
    The spectral extraction region is enclosed by the cyan line.
    The magenta line represents the background region, where we excluded a bright point source.
        }
    \label{fig:bandimage}
\end{figure}

\subsection{X-ray Spectroscopy}\label{sec:X}
Figure~\ref{fig:bandimage} shows the background-subtracted X-ray image of G359.0$-$0.9.
We found that only the eastern part is bright in X-ray within the radio shell whereas the other region is comparable with the background level outside of the remnant.
It is also notable that the X-ray surface brightness  is rather uniform in the east and that no apparent shell-like structure is seen there, which implies that G359.0$-$0.9 belongs to the class of Mixed-Morphology (MM) SNRs \citep{rho_1998}.
If this is the case, X-ray emission we detected in G359.0$-$0.9 is from shock-heated ejecta, as is the case with the other MM SNRs such as W44 \citep{uchida_2012} and  G359.1$-$0.5 \citep{suzuki_2020}.

\begin{figure}[!t]
    \centering
    \includegraphics[width=\linewidth]{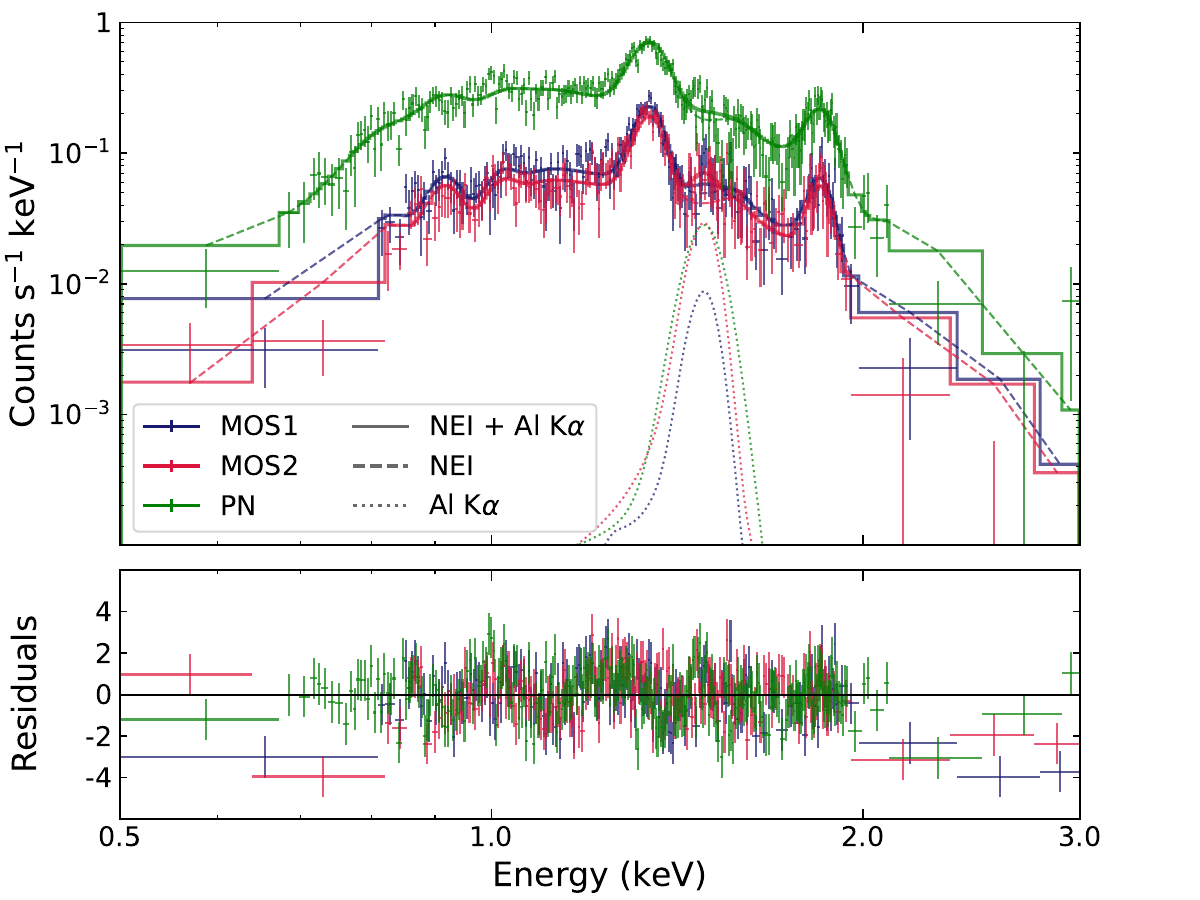}
    \caption{
    X-ray spectrum of G359.0$-$0.9 obtained with MOS1 (black), MOS2 (red), and pn (green).
    The best-fit model is represented by the dotted lines.
    The residuals are displayed in the lower panel.
    }
    \label{fig:bestfit}
\end{figure}

For the following  analysis, we focus on the bright eastern part of G359.0$-$0.9 as shown in Figure~\ref{fig:bandimage}.
The background region was chosen from the vicinity of the remnant, avoiding the point-spread function from nearby bright  binaries (SLX~1744–300/299).
Figure~\ref{fig:bestfit} displays the obtained spectrum, in which we detected prominent emission lines of highly ionized Mg and Si; no apparent line features of the Fe-L complex and Ne are found.
We applied an absorbed non-equilibrium ionization collisional plasma model (NEI; \texttt{vnei} in XSPEC), which is based on \citet{bamba_2009}.
We set the electron temperature $kT_e$, ionization timescale $n_et$, normalization and abundances of Ne, Mg, Si (=S=Ar=Ca), Fe (=Ni) as free parameters.
The column density $N_{\rm H}$ was also varied using \texttt{tbabs} \citep[T\"{u}bingen-Boulder absorption model;][]{wilms_2000}.
We also added a Gaussian component at 1.48~keV, which represents a known Al K$\alpha$ line emission due to the particle-induced background of EPIC \citep{carter_2007}.
The single-component NEI model provides a successful fit for the spectrum, as shown in Figure 2. 
Although residuals are seen especially around the $\sim1$~keV band, the widths of these humps are smaller than the response function of EPIC.
The addition of another NEI component or freeing the O abundance does not significantly improve the fit, hence we conclude that these residuals are attributed to statistical uncertainty.


The best-fit parameters are listed in Table~\ref{tab:bestfit_par}.
The result is almost consistent with the previous studies \citep{bamba_2000}, whereas we additionally measured abundances of Ne and Si for the first time.
Since Fe is less abundant than the other elements, which exceed the solar value, the plasma is most likely ejecta and thus G359.0$-$0.9 would be a remnant of a core-collapse SN.
The most striking result of our analysis is the high Mg abundance against Ne; $Z_{\rm Mg}/Z_{\rm Ne}=(\m{Mg}/\m{Ne})/(\m{Mg}/\m{Ne})_\odot \sim 1.90^{+0.27}_{-0.19}$ (mass ratio  $\sim 0.66^{+0.09}_{-0.07}$) deviates significantly from $Z_{\rm Mg}/Z_{\rm Ne}\sim1$ that is expected in most of SNRs.
We therefore conclude that G359.0$-$0.9 is another example of the Mg-rich SNRs \citep{park_2003a}, which has been reported for only two samples: N49B \citep{park_2017} and G284.3$-$1.8 \citep{williams_2015}. 
We additionally note that the Si abundance is also significantly higher than that of Ne and even more abundant than that of Mg; $Z_{\rm Si}/Z_{\rm Mg}=(\m{Si}/\m{Mg})/(\m{Si}/\m{Mg})_\odot \sim 1.30^{+0.16}_{-0.17}$ (mass ratio  $\sim 1.11\pm0.14$).
 Such a characteristic abundance pattern implies a somehow unusual nucleosynthesis during the stellar evolution or explosion.

\begin{table}[!t]
    \centering
    \label{tab:bestfit_par}
    \caption{The best-fit Parameters.
    }
    \begin{tabular}{lll}
    \hline
        Components & Parameters & Best-fit values \\
    \hline\hline
       Absorption (TBabs) & $N_{\rm H}~[10^{22}\m{cm}^{-2}]$  & $2.13^{+0.08}_{-0.05}$ \\
        NEI & $kT_e$~[keV] & $0.2715^{+0.0009}_{-0.0008}$ \\
            & $Z_\m{Ne}$ & $1.02^{+0.19}_{-0.20}$ \\
            & $Z_\m{Mg}$ & $1.93^{+0.19}_{-0.17}$ \\
            & $Z_\m{Si}=Z_\m{S}=Z_\m{Ar}=Z_\m{Ca}$ & $2.52^{+0.16}_{-0.19}$ \\
            & $Z_\m{Fe}=Z_\m{Ni}$ & $0.60^{+0.26}_{-0.15}$ \\
            & $Z_\m{other}$ & 1~(fix) \\
            & $n_et ~[\m{cm^{-3}s}]$ & $>10^{13}$\\
            & $\m{norm}~[10^{-2}]$ & $5.5^{+0.1}_{-0.3}$\\
        Gaussian    & E~[keV]  & 1.48~(fixed) \\
            & $\m{norm_{MOS1}}~[10^{-4}]$ & $8.6^{+8.1}_{-7.7} $ \\
            & $\m{norm_{MOS2}}~[10^{-3}]$ & $2.8^{+0.7}_{-0.6} $ \\
            & $\m{norm_{pn}}~[10^{-3}]$ & $3.4^{+1.8}_{-1.6} $ \\
    \hline
        & $\chi^2/\m{d.o.f}$ & $1665.5/1492$\\
    \hline
    \end{tabular}
\end{table}

\subsection{Discovery of a molecular cloud associated with G359.0$-$0.9}\label{cloud}

\begin{figure*}[t]
\begin{center}
\includegraphics[width=140mm]{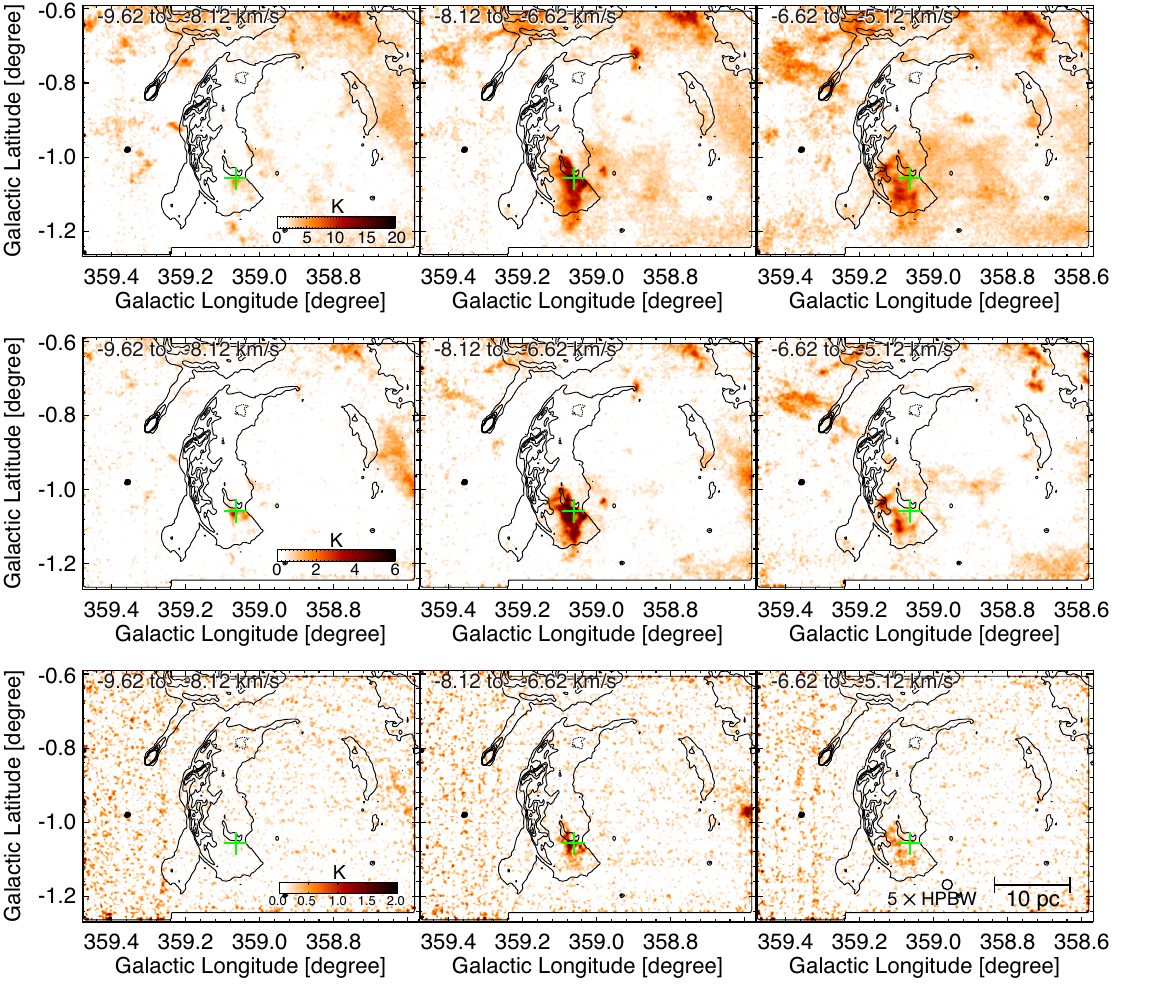} 
\caption{Velocity channel distributions of $\twelvecol$ (top three panels), $\thirteencol$ (middle three panels), and $\ceighteenol$ (bottom three panels) with smoothed 1.3~GHz contours obtained by MeerKAT. The 10~pc scale bar at 2.5~kpc and the HPBW are indicated in the bottom right panel. The light green cross indicates the position at which the molecular line spectra shown in Figure~$\ref{spec}$ have been obtained.}
\label{LBch}
\end{center}
\end{figure*}

\begin{figure}[t]
\begin{center}
\includegraphics[width=7cm]{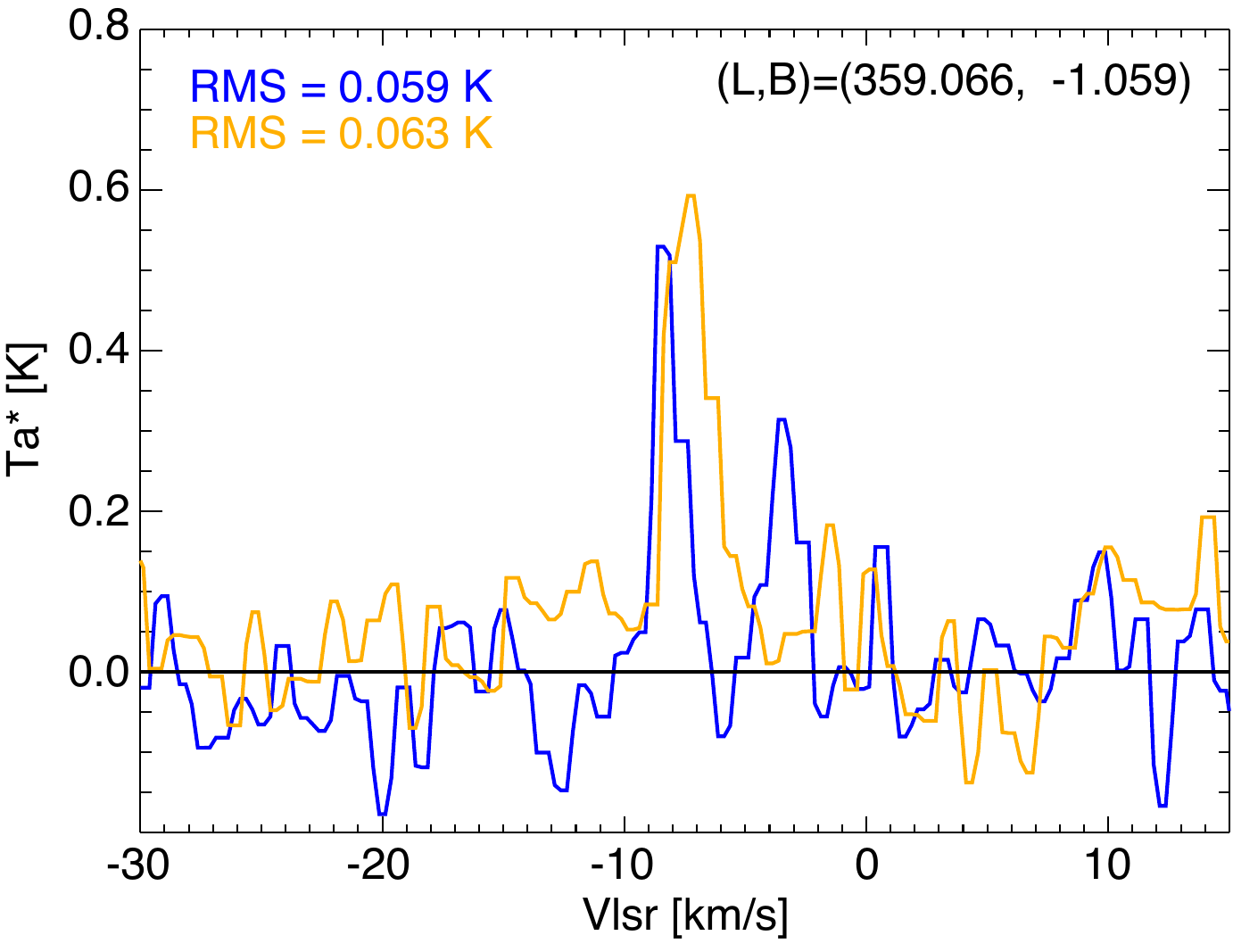}
\caption{
Molecular line spectra of $\hcnl$ (blue) and $\hcopl$ (orange) toward ($l$, $b$) = (359\fdg 066, $-$1\fdg 059). R.M.S. noise intensities in $\tastar$ are indicated at the top left of the panel.
}
\label{spec}
\end{center}
\end{figure}

Although the investigation for interactions with surrounding material is important for measuring the distance, no evidence has been reported for G359.0$-$0.9 so far.
Figure~\ref{LBch} shows velocity channel distributions of $\twelvecol$, $\thirteencol$, and $\ceighteenol$ obtained with our new observations. 
The optically thick $\twelvecol$ emission prevails almost over the entire observation field, while $\thirteencol$ and $\ceighteenol$ are detected only toward the bright parts in $\twelvecol$.
We discovered a bright CO clump seen in all three CO isotopes toward ($l$, $b$) = (359\fdg00 to 359\fdg12, $-$1\fdg22 to $-$1\fdg02).
The cloud displays complex structure with a sharp edge at the boundary of the remnant and coincides with the edges of  the southeastern radio-bright rims, which  suggests an interaction between the cloud and the SN shock.
As displayed in  Figure~\ref{spec},  the spectra obtained by
our follow-up observations under setting~2 (see Section~\ref{sec:nobeyama}) clearly show $\hcnl$ and $\hcopl$ lines from the cloud with a sufficient significance ($\sim$9--10$\sigma$).
According to \citet{set04}, HCO$^+$ lines trace collisionally excited, higher density spots owing to their large electric dipole moment, and they are mainly detected in SN shock regions. Thus, we conclude that the southeastern dense cloud at $\vlsr \sim$$-$7~$\kms$ is associated with G359.0$-$0.9.
The diffuse envelope of the associated cloud may be blown away by the stellar wind from the progenitor, and the surviving dense part is interacting with the remnant.

\section{Discussion}\label{sec:dis}

\subsection{Origin of G359.0$-$0.9 and Mg-rich SNRs}

\begin{figure*}[!t]
        \centering
        \includegraphics[width=0.6\linewidth]{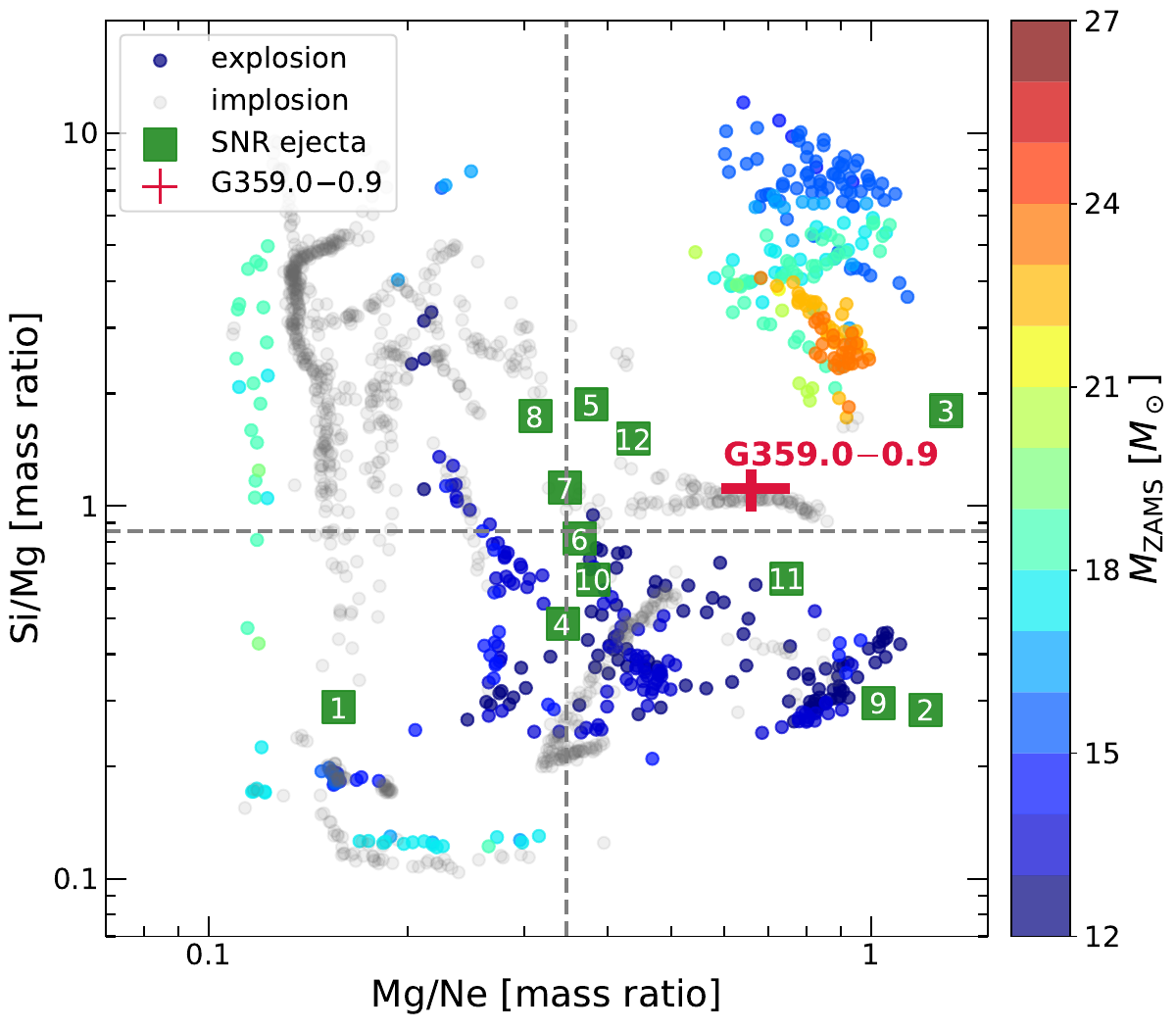}
        \caption{
        Parameter distribution of Mg/Ne and Si/Mg expected from stellar models \citep[circles;][]{sukhbold_2018} and obtained from previous observations of SNRs (green squares).
        The colored and gray circles represent explosion and implosion cases, respectively.
         The result of G359.0$-$0.9 is shown by the red cross.
         The gray dashed lines represent the solar values.
         The numbers correspond to the targets as follows;         
        [1] 1E~0102.2$-$7219; \citet{sasaki_2001}, 
        [2] G284.3$-$1.8; \citet{williams_2015}, 
        [3] G290.1$-$0.8; \citet{kamitsukasa_2015}, 
        [4] MSH~15$-$5\textit{2}; \citet{yatsu_2005}, 
        [5] MSH~15$-$5\textit{6}; \citet{yatsu_2013}, 
        [6] N132D; \citet{hughes_1998}, 
        [7] N23; \citet{uchida_2015}, 
        [8] N49; \citet{uchida_2015},
        [9] N49B; \citet{uchida_2015},
        [10] N63A; \citet{hughes_1998},
        [11] RX~J1713.7$-$3946; \citet{katsuda_2015}, 
        [12] W44; \citet{uchida_2012}.
        }  
        \label{fig:MgNe_SiMg}
\end{figure*}

\begin{figure*}[t]
        \centering
        \includegraphics[width=0.9\textwidth]{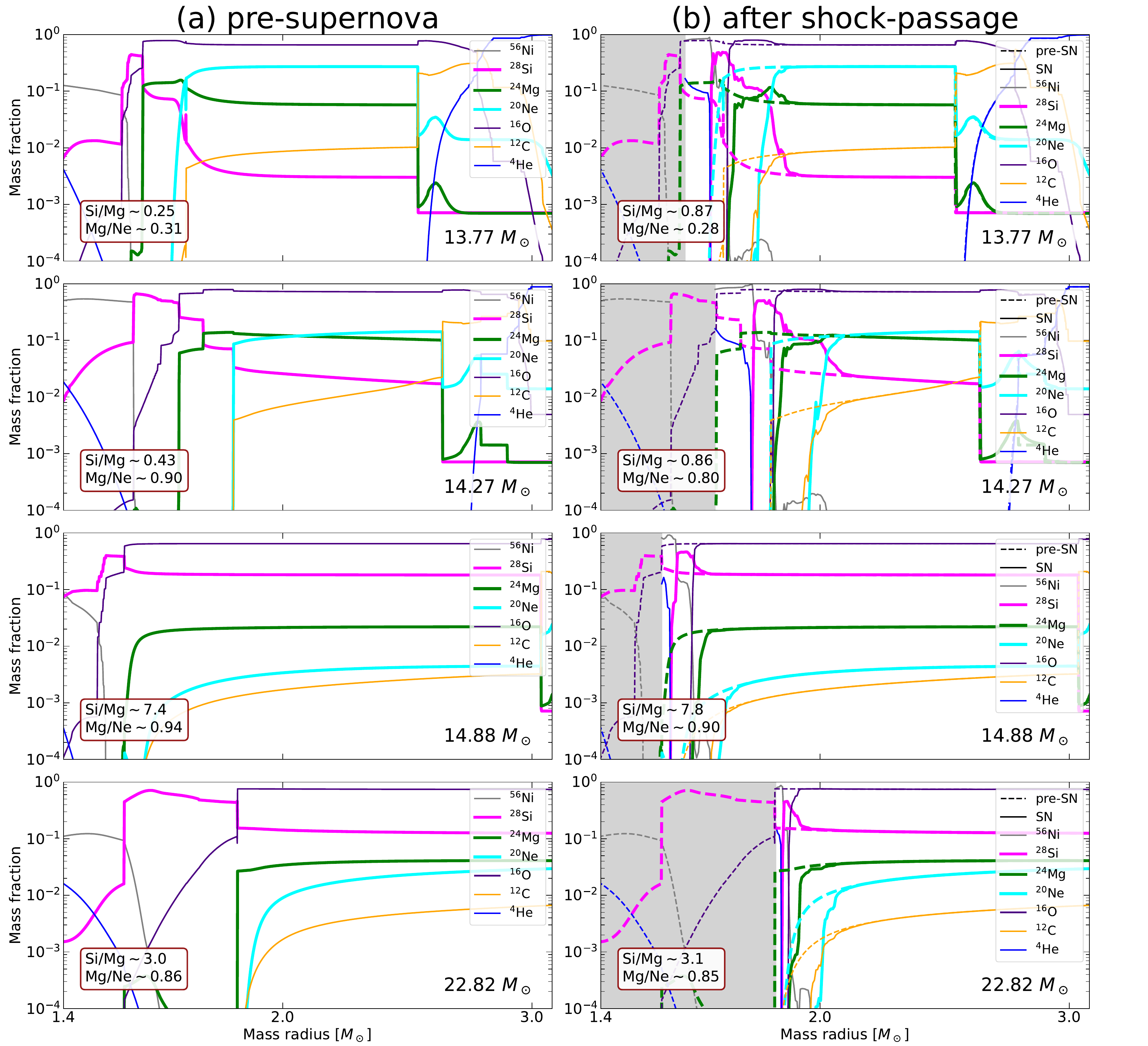}
        \caption{
        Mass fraction profiles of stellar models in pre-SN \citep[left;][]{sukhbold_2018} and after shock passage (right).
        The colors of lines represent each element.
        From the top to the bottom panels, we show the cases of $M_{\rm ZAMS}=13.77M_\odot$, $14.27M_\odot$, $14.88M_\odot$ and $22.82M_\odot$.
        The inset indicates  expected mass ratios of Si/Mg and Mg/Ne, which are obtained by integrating elements in the O-rich and the outer layers.
        The grey shaded area represents a  mass radius of a PNS.
        } 
      \label{fig:massfraction}
\end{figure*}

As stated in Section~\ref{sec:X}, our spectral analysis of G359.0$-$0.9 indicates that the Mg (and also Si) abundance is significantly higher than the solar value when compared with that of Ne.
We have categorized this remnant as a Mg-rich SNR defined by \citet{park_2003a}.
While only two SNRs have been identified as Mg-rich SNRs so far \citep{park_2017, williams_2015}, we expect more candidates to be identified as is the case with G359.0$-$0.9.
Based on a literature search, we found that at least two SNRs, RX~J1713.7$-$3946 \citep{katsuda_2015} and G290.1$-$0.8 \citep{kamitsukasa_2015}, may possibly fall into the ``Mg-rich''category.
In Figure~\ref{fig:MgNe_SiMg}, we plot the abundance ratios among Ne, Si, and Mg obtained in our spectral fit of G359.0$-$0.9 and previous measurements of other luminous SNRs.
The result supports that  several SNRs including G359.0$-$0.9 show higher Mg/Ne ($\sim0.66^{+0.09}_{-0.07}$) while the others are roughly clustered around  the solar value ($\sim0.35$).

We recently proposed that the abundant Mg observed in N49B is likely due to a destratification in the progenitor \citep{sato_2024a}, where the O- or Ne-burning shell is breaking into outer layers shortly before the core collapse \citep[shell merger;][and references therein]{yadav_2020}.
Such violent destratification promotes a series of Ne-burning processes like $\m{^{20}Ne(\alpha,\gamma)^{24}Mg}$ and in the case of an O-burning shell merger, $\m{^{20}Ne(\alpha,\gamma)^{24}Mg(\alpha, \gamma)^{28}Si}$ is further enhanced.
This scenario can account for the  origin of the Mg-rich SNRs and we speculate that the abundance ratios among Ne, Si, and Mg are  the keys to deciphering stellar nucleosynthesis and evolution.
The Ne-burning processes before the explosion result in a higher Mg/Ne, i.e., Mg-rich ejecta, whereas if the O-burning shell merger undergoes, the resultant Si/Mg also becomes higher.
Our spectral analysis of G359.0$-$0.9 shows a relatively high Si/Mg mass ratio $\sim1.11\pm0.14$ compared with those of N49B and G284.3$-$1.8 \citep[$\sim$0.3][]{park_2017, williams_2015}, which may be explained by the shell merger with/without the O-burning shell merger.

In order to compare our results with theoretical expectations, we calculate a nucleosynthesis model taking into account the destratification.
We used 1D stellar simulations with a standard mass-loss rate published by \citet{sukhbold_2018}.
The results are overlaid in Figure~\ref{fig:MgNe_SiMg}, where we distinguish whether the models are exploding or non-exploding by evaluating a threshold for the explodability (computed with the W18 engine) according to \citet{ertl_2016}.
Note that we calculated the values of Mg/Ne and Si/Mg outside than the O-burning shell, where the elemental composition is not changed drastically before/after a shock passage.
Since Mg and Ne are mainly  synthesized before core collapse whereas Si is enriched also in the explosive nucleosynthesis,  the calculated values and the measured abundances (particularly of Si) should not be directly compared in Figure~\ref{fig:MgNe_SiMg}.  
 It would be nevertheless remarkable that the calculated trend  roughly coincides with the observations and especially that there are two ``Mg-rich'' groups with different Si/Mg.
We found that the Mg-rich group with a high Si/Mg ratio indicates a relatively high $M_\m{ZAMS}$ ($>14M_\odot$), and it is particularly notable that all the results of massive progenitors with $M_\m{ZAMS}$ $>20M_\odot$ are distributed only in this group.
The result suggests that the high Si/Mg ratio observed in the ejecta of a Mg-rich SNR is a good indicator of a remnant of a massive progenitor: only a SNR that exhibits both Mg-rich ejecta and a high Si/Mg ratio would be a plausible candidate for a massive star's remnant ($M_\m{ZAMS}>20M_\odot$).

\subsection{Explosive Nucleosynthesis Yield}

\begin{figure}[t]
    \centering
    \includegraphics[width=0.91\linewidth]{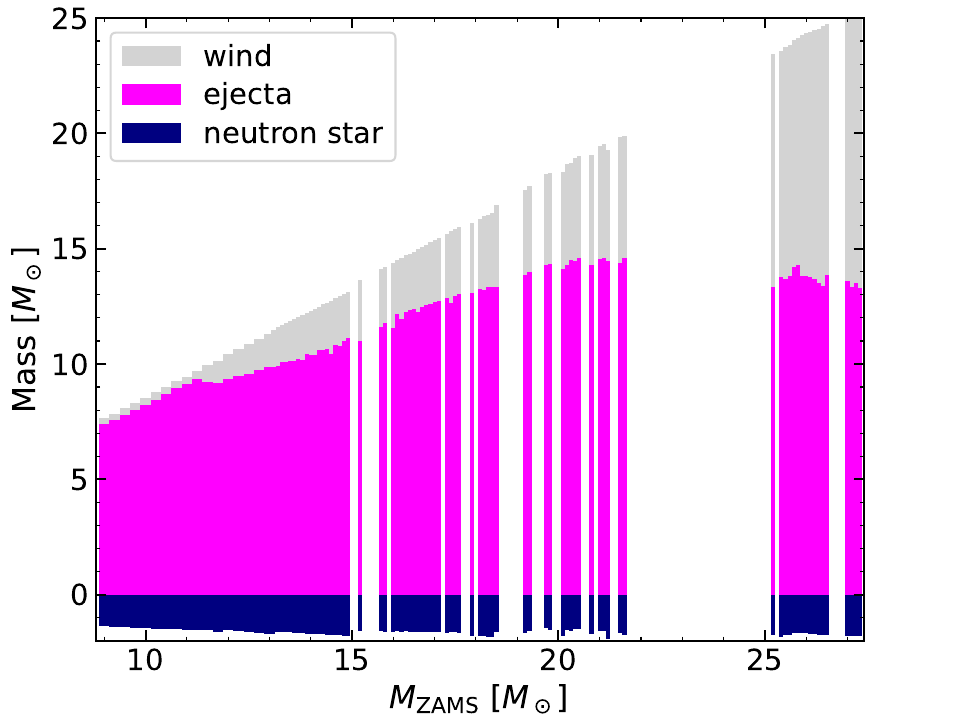}
    \caption{
    Mass distributions of  a compact object (dark blue), ejecta (magenta), and stellar wind (grey) for different $M_\m{ZAMS}$ based on the dataset of the Z9.6 and W18 engines given by \citet{sukhbold_2016}.  
    }
    \label{fig:suh_ejmass}
\end{figure}

As displayed in the left panels of Figure~\ref{fig:massfraction}, 
we confirmed 
that the shell merger triggers efficient Ne-burning processes and enhances the total yield of Mg and Si, on the basis of the stellar simulations by \citet{sukhbold_2018}.
In the case of the 13.77$M_\odot$ progenitor, for instance, the O-Mg layer are formed as a result of Ne-burning processes, keeping up the boundary with the outer O-Ne layer.
On the other hand, due to the shell merger, the boundary in the 14.27$M_\odot$ progenitor seems to disappear \citep[Ne-burning shell intrusion; cf.][]{sato_2024a} and also in the 14.88$M_\odot$ and 22.82$M_\odot$ progenitors O-Si layers are formed by mixing the O-Ne or O-Mg layers \citep[O-burning shell merger; cf.][]{sato_2024a}.
We found that these merger processes tend to enhance and reduce the total amount of Mg and Ne, respectively.
Our simulation reveals that all the models  whose Mg/Ne ratio is higher than $\sim0.6$ can be explained by such destratification.
It can be used as a threshold to determine whether a SNR is Mg-rich or not.

In order to check how the subsequent nucleosynthesis affects the final yields of Ne, Mg, and Si, we run 1D SN simulations using The SuperNova Explosion Code \citep[SNEC,][]{morozova_2015} with a Lagrangian hydrodynamic code, in which we incorporate neutrino heating and cooling via a light-bulb scheme \citep[][]{suwa_2019}.
The right panels of Figure~\ref{fig:massfraction} displays the resultant mass fraction profiles for various $M_\m{ZAMS}$ after shock passage.
We found no significant change in Mg/Ne although a slight decrease is seen: the aforementioned threshold $\sim0.6$ is still able to discriminate between Mg-rich  and normal SNRs.
While the explosive nucleosynthesis gives an overall trend of increase in Si/Mg, only a slight increase is seen in the cases of the O-burning shell merger, i.e., $M_\m{ZAMS}=14.88M_\odot$ and 22.82$M_\odot$.
This may be because before core collapse the O-burning shell merger produces a large amount of Si  in the O-rich layer, where the explosive nucleosynthesis is less effective.
It is remarkable that all massive star models ($>20M_\odot$) experience the O-burning shell merger and hence the resultant Si/Mg becomes significantly higher than the solar value \citep[0.86;][]{wilms_2000}.
We therefore believe that if a Mg-rich SNR exhibits high Si/Mg as well it may  possibly be a remnant of a massive star.
In this sense, since Si/Mg in G359.0$-$0.9 ($\sim1.11\pm0.14$) is close to the solar value and relatively consistent with the 14.27$M_\odot$ case, the progenitor is likely smaller than 15$M_\odot$ and was experienced the Ne-burning shell intrusion before the explosion.

\begin{figure}[t]
    \centering
    \includegraphics[width=0.91\linewidth]{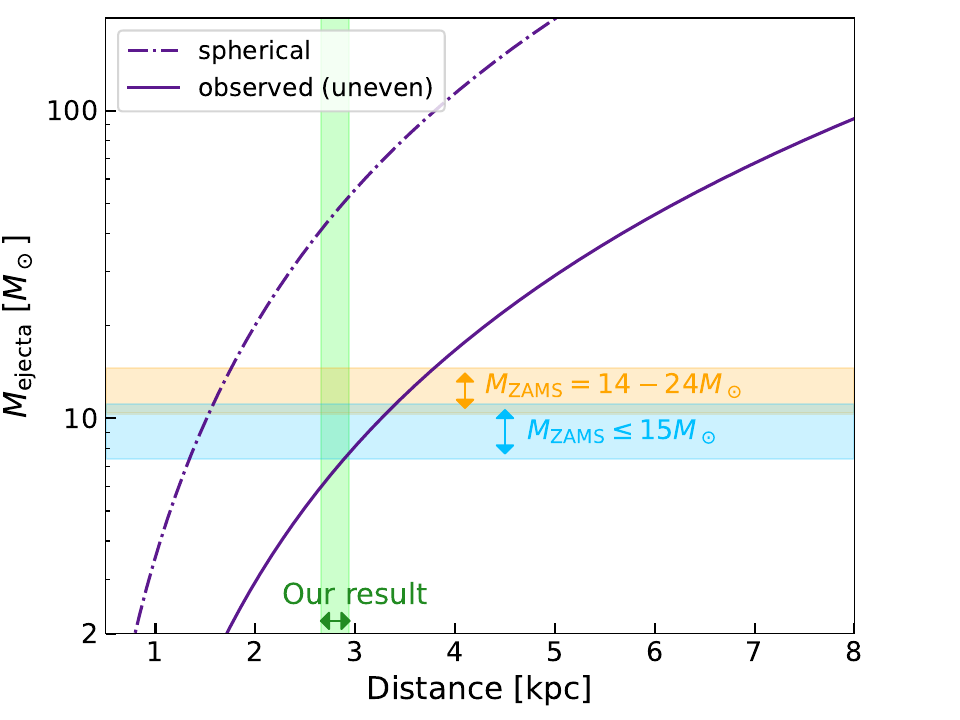}
    \caption{Expected $M_{\rm ejecta}$ of G359.0$-$0.9 versus distance plot.
    The solid and dash-dotted lines indicate the results assuming the uneven (aspherical) and completely spherical ejecta distribution, respectively.
    The green hatched area represents the accepted distance range determined by our radio observation with NRO 45-m.
    The orange and blue hatched regions show   ejecta mass ranges of 14--24$M_\odot$ and 12--15$M_\odot$, which are expected for the Mg-rich SNRs (Figure~\ref{fig:suh_ejmass}).
%
    }
    \label{fig:md}
\end{figure}

\subsection{Estimation of  the ejecta mass of G359.0$-$0.9}

As noted in the previous section,  our nucleosynthesis calculations prefer a progenitor mass ($M_\m{ZAMS}$) less than 15$M_\odot$ as the origin of G359.0$-$0.9, which should be consistent with the ejecta mass estimate based on the observation.
Figure~\ref{fig:suh_ejmass} shows the relationship between $M_\m{ZAMS}$ and the ejecta mass on the basis of calculations given by \citet{sukhbold_2016}.
The result suggests that the ejecta mass of G359.0$-$0.9 is expected to be less than $\sim10M_\odot$ in total.
In order to measure the ejecta mass from the emitting volume of  this remnant, we require an accurate  distance measurement.
Nevertheless, whether G359.0$-$0.9 is a nearby SNR or not is still a question in the previous studies;
$3.5\pm{0.4}~\m{kpc}$ with UKIDSS and $3.3\pm{0.2}~\m{kpc}$ with VVV \citep[luminosity attenuation of red clumps][]{wang_2020}, $\sim3.7~\m{kpc}$ \citep[the $\Sigma$-$D$ relation;][]{pavlovic_2012}, and $\sim6~\m{kpc}$ \citep[X-ray absorption;][]{bamba_2000}.

We estimate the distance to G359.0$-$0.9 on the basis of the newly discovered cloud (Section~\ref{cloud}) using a similar method applied for G359.1$-$0.5 by \citet{suzuki_2020}.
This is one of the most reliable methods to obtain the distance to a SNR \citep[cf.][]{fukui_2012}. 
The associated cloud has a small velocity width $\sim2$~$\kms$, and its velocity is very close to 0~$\kms$. 
These properties strongly suggest its location in a foreground spiral arm rather than in the Galactic Center \citep[see][for details]{eno14, eno23}.
According to a spiral arm model of our Galaxy \citep[e.g.,][]{reid_2016} , the velocity of the associated cloud well coincides with that of the Scutum-Centaurus arm.
We thus conclude that the associated cloud is likely located in the Scutum-Centaurus arm.
Based on Figure 5 in \citet{eno23}, we estimated the distance of the Scutum-Centaurus arm in the direction of G359.0$-$0.9 to be 2.8 kpc.
Since the width of the Scutum-Centaurus arm is obtained to be  $\sim0.14$~kpc \citep[][see Figure~5]{rei16},  the distance $d$ to G359.0$-$0.9 is estimated to be between 2.66~kpc and 2.94~kpc.
The real radius $r$ of the remnant in this case  is $\sim11$~pc.


Since G359.0$-$0.9 exhibits a highly asymmetric morphology, we  consider two cases with different 3D structures for calculating the entire volume: the ejecta has an intrinsically uneven distribution due to an aspherical explosion or a completely spherical  with $r=11$~pc but emitting region is somehow limited in the east.
In the former case, we assume a spherical cap whose opening solid angle is $\pi/2$, which is consistent with a previous estimate \citep{bamba_2000}.
The ejecta mass is then calculated as follows:
\begin{equation}
M_{\rm ejecta}~\sim 6.8 \left(\frac{d}{2.8\m{~kpc}}\right) \left(\frac{f}{0.9}\right) \left(\frac{\rm{norm}}{5.5\times10^{-2}~\rm{cm}^{-5}}\right) M_\odot,
\end{equation}
where the volume filling factor $f$ was deduced from a fluctuation in X-ray surface brightness.
The result of our  calculation is displayed in Figure~\ref{fig:md}, where  the latter spherical case  overlaid as well.
We found that $M_{\rm ejecta}$ only assuming the latter uneven case lies within an acceptable range given by \citet{sukhbold_2016} and that on the contrary the former spherical case requires much larger $M_{\rm ejecta}$ than theoretically expected.
Given that the ejecta of G359.0$-$0.9 is unevenly distributed to the east, the progenitor mass $M_\m{ZAMS}$ seems to be less than 15$M_\odot$, following our expectation from the abundance pattern.

\section{Conclusions}
We performed detailed observations of a Galactic SNR G359.0$-$0.9 in X-ray and radio bands, which reveal a distorted X-ray morphology with a shell-like structure.
The result implies that this remnant belongs to the class of mixed-morphology SNRs.
Our spectral analysis with XMM-Newton allows us to measure accurate ejecta mass ratios of Mg/Ne ($\sim0.66^{+0.09}_{-0.07}$; solar value is $\sim0.35$) and Si/Mg  ($\sim1.11^{+0.14}_{-0.14}$) for the first time.
The result indicates that this remnant is categorized as the Mg-rich SNRs \citep{park_2003a}, whose origin is still under debate.
Our conclusions are presented below:

\begin{enumerate}
\item On the basis of our ongoing study \citep{sato_2024a}, we speculate that such Mg-rich abundance pattern is derived from violent destratification, so-called shell merger processes that are expected to occur before core collapse \citep[e.g.,][]{yadav_2020}.
By comparing stellar evolution models given by \citet{sukhbold_2018} and  our SN simulations  with the abundance ratios of previously observed SNRs,  we found that if we take $\rm{Mg/Ne}\sim0.6$ as the threshold several SNRs including G359.0$-$0.9  are labeled as the Mg-rich SNRs.
\item Our calculations  also reveal that there are two ``Mg-rich'' groups with different Si/Mg.
This is because there are two types of the destratification,  the Ne-burning shell intrusion and the O-burning shell merger, and the latter process enhances the total yield of Si.
Since all massive star models ($>20M_\odot$) experience the O-burning shell merger, a Mg-rich SNR with high Si/Mg may  possibly be evidence for a remnant of a massive star.
In this sense, the abundance pattern of G359.0$-$0.9 prefers the Ne-burning shell intrusion and thus we conclude that the progenitor mass $M_\m{ZAMS}$ is likely smaller than 15$M_\odot$.
\item We also discovered a  molecular cloud interacting with G359.0$-$0.9 with the Nobeyama 45~m telescope. 
Consequently, the distance ambiguity  that has been argued was solved in favor of the foreground remnant.
We conclude that G359.0$-$0.9 is  located in the Scutum-Centaurus arm, whose distance is $\sim2.8$~kpc (2.66--2.94~kpc).
Applying this result, we measured the total ejecta mass of $M_{\rm ejecta}\sim6.8M_\odot$, which is consistent with the above estimation of  $M_\m{ZAMS}<15M_\odot$.
Our result also indicates that the ejecta distribution is not spherical but highly distorted, which implies an asymmetric  SN explosion.
\end{enumerate}
%

\begin{acknowledgments}

This work is supported by JSPS/MEXT Science Research grant Nos. JP24KJ1485 (K.M.), JP19K03915, JP22H01265 (H.U.) and JP23KJ1350 (T.N.), and JSPS Core-to-Core Program grant No.~JPJSCCA20220002 (T.N.).
Based on observations obtained with XMM-Newton, an ESA science mission with instruments and contributions directly funded by ESA Member States and NASA.
The Nobeyama 45-m radio telescope is operated by Nobeyama Radio Observatory, a branch of National Astronomical Observatory of Japan.
We also thank M., Matsuda and S., Inoue for helpful advice in our analysis.
\facilities{XMM, NRO}
\software{XSPEC, HEASoft, SNEC}
\end{acknowledgments}

\clearpage
\bibliographystyle{aasjournal}
\bibliography{main}

\end{document}